\documentclass[twocolumn]{aastex631}
\usepackage{graphicx}
\usepackage{threeparttable}
\usepackage{amsmath}
\usepackage{bm}
\usepackage{booktabs}
\usepackage{multirow}

\newcommand{\FIG}[1] {Figure~\ref{#1}}
\newcommand{\TAB}[1] {Table~\ref{#1}}

\newcommand{\PMO}{\affiliation{Purple Mountain Observatory, Chinese Academy of Sciences, Nanjing 210023, P.~R.~China}}
\newcommand{\USTC}{\affiliation{School of Astronomy and Space Sciences, University of Science and Technology of China, Hefei 230026, P.~R.~China}}
\newcommand{\NJUPT}{\affiliation{School of Computer, Nanjing University of Posts and Telecommunication, Nanjing 210023, P.~R.~China}}
\newcommand{\UCASNJ}{\affiliation{University of Chinese Academy of Sciences, Nanjing 211135, P.~R.~China}}
\newcommand{\IHEP}{\affiliation{Key Laboratory of Particle Astrophysics, Institute of High Energy Physics, Chinese Academy of Sciences, Beijing 100049, P.~R.~China}}
\newcommand{\GXLAB}{\affiliation{Guangxi Key Laboratory for Relativistic Astrophysics, Nanning 530004, P.~R.~China}}
\newcommand{\SHAO}{\affiliation{Shanghai Astronomical Observatory, Chinese Academy of Sciences, 80 Nandan Road, Shanghai 200030, P.~R.~China}}
\newcommand{\AHNU}{\affiliation{Department of Physics, Anhui Normal University, Wuhu, Anhui 241002, P.~R.~China}}
\newcommand{\NJU}{\affiliation{School of Astronomy and Space Science, Nanjing University, Nanjing 210023, P.~R.~China}}
\newcommand{\JXSTU}{\affiliation{Department of Physics, Jiangxi Science \& Technology Normal University, Nanchang 330013, P.~R.~China}}
\newcommand{\NJULAB}{\affiliation{Key Laboratory of Modern Astronomy and Astrophysics (Nanjing University), Ministry of Education, Nanjing 210023, P.~R.~China}}
\newcommand{\KLRAT}{\affiliation{Key Laboratory of Radio Astronomy and Technology, Chinese Academy of Sciences, A20 Datun Road, Beijing 100101, P.~R.~China}}

\shorttitle{GRB timing}
\shortauthors{Geng et al.}

\begin{document}

\title{GRB timing: decoding the hidden slow jets in GRB 060729}

\author[0000-0001-9648-7295]{Jin-Jun~Geng}\thanks{E-mail: jjgeng@pmo.ac.cn}
\PMO
\author[0009-0002-7851-3706]{Ding-Fang Hu}
\PMO
\USTC
\author[0000-0001-7892-9790]{Hao-Xuan Gao}
\PMO
\author[0009-0005-0170-192X]{Yi-Fang Liang} 
\PMO
\USTC
\author{Yan-Long Hua}
\PMO
\USTC
\author{Guo-Rui Zhang}
\NJUPT
\UCASNJ
\author[0000-0003-1166-3814]{Tian-Rui Sun}
\PMO
\author[0000-0002-0238-834X]{Bing Li}
\IHEP
\GXLAB
\author[0000-0001-9321-6000]{Yuan-Qi Liu}
\SHAO
\author[0000-0001-7943-4685]{Fan Xu}
\AHNU
\author[0000-0002-2191-7286]{Chen Deng}
\NJU
\author[0000-0002-5238-8997]{Chen-Ran Hu}  
\NJU
\author{Ming Xu}
\JXSTU
\author[0000-0001-7199-2906]{Yong-Feng Huang}
\NJU
\NJULAB
\author[0000-0002-6388-649X]{Miao-Miao Zhang}
\PMO
\author[0000-0001-8060-1321]{Min Fang}
\PMO
\author[0000-0003-2553-2217]{Jing-Zhi Yan}
\PMO
\author[0000-0003-4341-0029]{Tao An}
\SHAO
\KLRAT
\author[0000-0002-6299-1263]{Xue-Feng Wu}\thanks{E-mail: xfwu@pmo.ac.cn}

\begin{abstract}
Gamma-ray bursts (GRBs) are luminous stellar explosions characterized by the ejection of relativistic jets. This work proposes a novel paradigm to study these GRB jets. By analyzing the timing information of prompt pulses and X-ray flares, in conjunction with the multi-wavelength afterglow observations,
we identify three distinct jets in the extraordinary GRB 060729, with initial bulk Lorentz factors ranging from approximately 20 to 80, smaller than typical values of $> 100$. These three jets undergo two successive collisions, producing the observed pair of X-ray flares.
Following these interactions, the system evolves into a fast, narrow jet and a slower, hollow jet that continues to propagate in the circumburst medium, evidenced by the notable twin bumps observed in the X-ray and optical afterglow of GRB 060729.
Our findings demonstrate that the timing of the early emission enables us to measure the velocities of the GRB jets.
The proposed paradigm enhances our understanding of jet dynamics and shock interactions and serves as a powerful tool for probing the physics of the central engine with the expanded sample in the current golden era of GRB research.
\end{abstract}

\section{Introduction}\label{sec:intro}
Gamma-ray bursts (GRBs) are intense flashes of gamma rays associated with catastrophic stellar explosive events, where relativistic jets are launched from the central engine~\citep{Piran99,Kumar15,Zhang18}.
As the variability in the engine's accretion process leads to the ejection of jet elements at different velocities, the dissipation of these elements through mechanisms such as internal shocks~\citep[e.g.,][]{Rees94} or magnetic reconnection~\citep{Zhang11} results in the prompt emission. 
After the erratic prompt phase, these outflows coalesce into an external shock that surfs in the circumburst environment, producing long-term multi-wavelength afterglows~\citep[e.g.,][]{Blandford76,Sari99,Huang00,WangXG2015}.
However, extensive observational data has revealed a variety of temporal behaviors that challenge the afterglow evolution patterns predicted by the standard model.
Phenomena such as flares, plateaus, and re-brightenings in the X-ray ~\citep{Gehrels04,Burrows05,Zhang06,Tang19} and optical bands~\citep{Liliang12,Kann24} suggest the involvement of complex physical processes, including multiple jet components~\citep{Berger03,Sheth03,Huang04,Peng05}, complex shock physics~\citep{Kong10}, and prolonged activity of the central engine~\citep{Kumar08,Wu13} at the late stages. 

The X-ray plateau, in particular, has inspired several theoretical models. These include scenarios involving energy injection processes~\citep{Dai98,Dai04,Fan06,Yu07,Xu09,Geng13,vanEerten14,Geng16}, the refreshing of the forward shock by late-arriving shells~\citep{Zhang01}, the structured jet viewed off axis~\citep{Beniamini20}, and a slow jet moving through a wind environment~\citep{Begue22}.
A significant advancement in our understanding came from the observation of GRB 240529A, a double-yolk burst featuring two energetic gamma-ray pulses separated by approximately 400 seconds~\citep{Konus-Wind}. 
While the afterglow of this burst displayed a typical X-ray plateau and optical re-brightening, detailed synergistic analyses of both the prompt and the afterglow emission suggest that these features result from two distinct shocks launched separately from the central engine~\citep{Sun24}.
Notably, the initial bulk Lorentz factor of the later shock in GRB 240529A was only about 50, significantly lower than the typical values of several hundred observed in other GRBs, thereby shedding light on the underlying physics of the central engine~\citep{Sun24,Lei13,Liu17}.
The peculiarities of GRB 240529A raise the question of whether it is an anomaly or indicative of a broader, previously overlooked phenomenon.
If multiple shocks are indeed launched from individual main pulses within one burst, it is anticipated that the interaction between shocks with varying properties — such as different initial jet opening angles and bulk Lorentz factors — could lead to complex physical phenomena. 
This scenario warrants further investigation to enhance our understanding of GRB dynamics and the mechanisms driving their diverse observational signatures.

In this Letter, we introduce a novel GRB synthesis paradigm designed to probe the properties of GRB jets by integrating all available information, rather than modeling different phases independently. While previous studies have utilized the onset features of afterglows to constrain the initial velocity of external shocks~\citep[e.g.,][]{Liang10}, the information from the early emission phase has often been overlooked. By analyzing the timing characteristics of prompt pulses and X-ray flares, combined with multi-wavelength afterglow observations, we achieve a deeper exploration
of the intricate physical processes that unfold during GRB events, particularly those involving multiple shocks with diverse properties. GRB 060729 serves as a prototype successfully decoded within this new paradigm.

This Letter is organized as follows: In Section 2, we review the observations of GRB 060729. The framework of GRB timing is present in Section 3. The numerical results and relevant implications are given in Sections 4 and 5. We summarize our results in Section 6. 

\section{Observations} \label{sec:obs}
\begin{figure*}[ht]
	\centering
	\includegraphics[width=\textwidth]{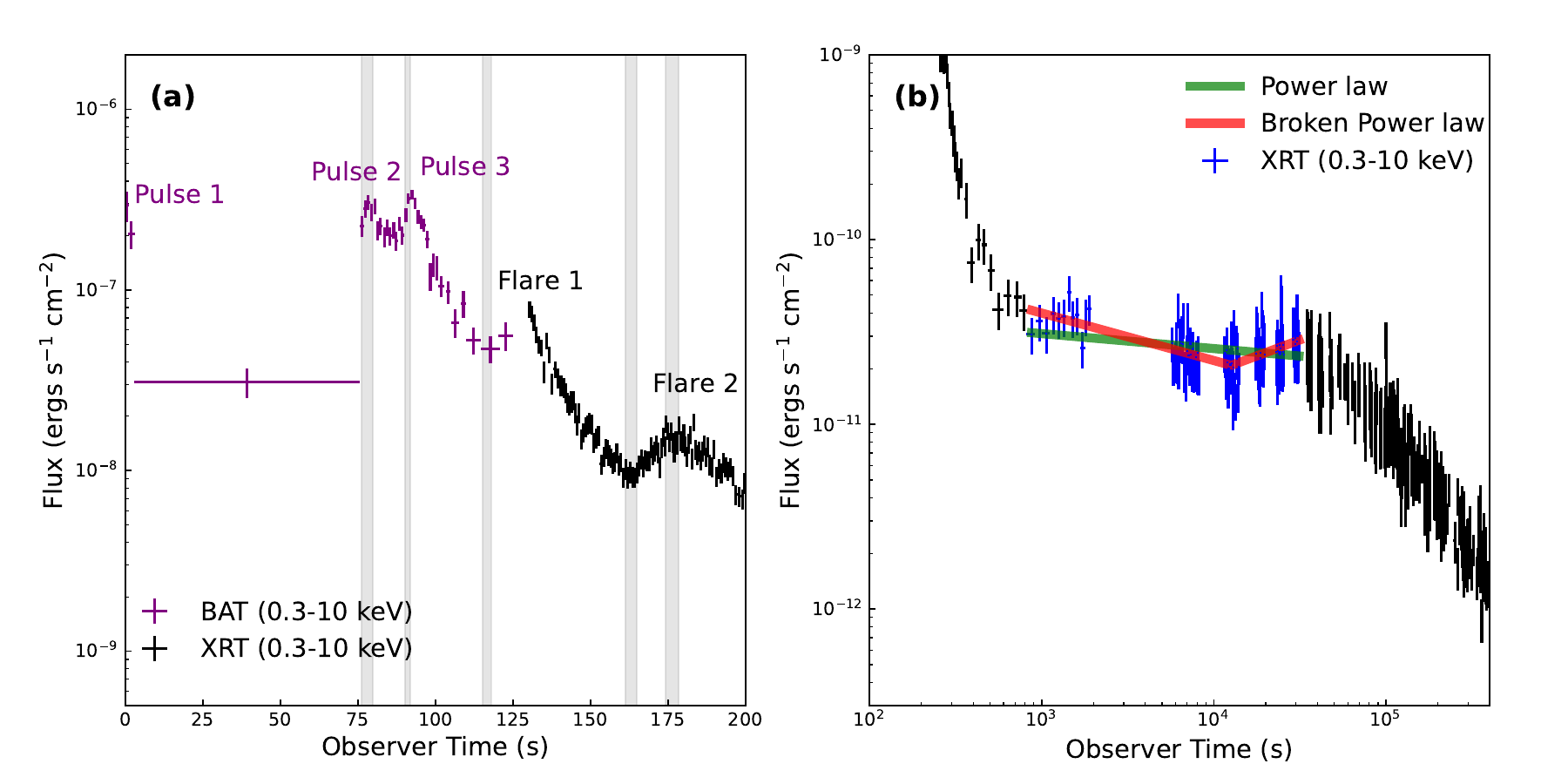}
	\caption{The X-ray lightcurve detected by {\it Swift}.
	Panel (a) shows the combined BAT (purple) and XRT (black) lightcurves within 200~s after the trigger, which consists of three pulses during the main burst and two flares.
	The vertical grey shaded areas mark several critical time ranges ($t_2^{\rm obs}$, $t_3^{\rm obs}$, $t_{\rm coll,1}$, $t_{\rm coll,2}$ and $T_{\rm peak}$) discussed in the main text.
	The BAT data are taken from \cite{Grupe07} and the XRT data are taken from the {\it Swift}/XRT website (http://www.swift.ac.uk/xrt\_curves/00221755/).
	Panel (b) exhibits the long-term evolution of X-ray emissions beyond 200~s by XRT (black points). A segment of the data marked in blue color thought as a ``plateau'' phase, is fitted by the power law (green line) and broken power law (red line), respectively.} \label{fig:XRT}
\end{figure*}

GRB 060729 is a bright burst discovered by the Neil Gehrels {\it Swift} Observatory ({\it Swift}; \citealt{Gehrels04,Burrows05}) on 2006 July 29
\citep{Grupe07}. The Burst Alert Telescope (BAT; \citealt{Barthelmy2005BAT}) onboard detected a main burst that lasts roughly 100 s and consists of three pulses, and the X-Ray Telescope (XRT; \citealt{Burrows2004XRT}) captures two subsequent X-ray flares that emerge within $\sim$ 100-200 s (\FIG{fig:XRT}).
The afterglow of this burst was exceptionally bright both in X-rays and UV/optical bands.
After a short steep decay phase, the X-ray emission transitions into a long-lasting plateau phase until $\sim 10^4$~s.
Meanwhile, GRB 060729 was well followed by the UV/Optical telescope (UVOT; \citealt{Roming2005UVOT}) up to 9 days post-burst in six UVOT filters,
showing remarkable similarities to the X-ray behavior, 
i.e., their temporal slopes and break times~\citep{Grupe07,Grupe10}.

While previous studies interpreted the X-ray lightcurve as a ``plateau'', the statistics analysis supports that it actually exhibits a synchronous increase with the UV/optical re-brightening around $10^4$~s at the 7$\sigma$ level.
In Figure 1, two analytical models, i.e., power-law and broken power-law, are adopted to fit the segment of X-ray data that is traditionally treated as a ``plateau''. The prevailing Akaike Information Criterion AIC = $2 K + \chi^2$ and Bayesian Information Criterion BIC = $K \ln (\mathcal{N}) + \chi^2$ are used to evaluate the fitting goodness, where $\chi^2$ is the Chi-square between the model and the data, $K$ and $\mathcal{N}$ are the numbers of model parameters and data respectively.
It gives AIC = 163.3 and BIC = 175.1 for the broken power-law fit to the plateau data versus AIC = 223.3 and BIC = 229.2 for the single power-law fit. Such difference in AIC and BIC suggests that the broken power-law is preferred at a $7\sigma$ level, which supports the re-brightening feature of the X-ray lightcurve around $\sim 10^4$ (see Appendix \ref{append:A} for further discussions).

The shallow decay behaviors in X-ray lightcurves are expected in the traditional energy injection scenarios, X-ray re-brightenings may further require that the injection luminosity is somehow increasing with time at $\sim 10^4$~s post burst.
Moreover, the UV afterglow suggests an early peak around $\sim 2 \times 10^3$~s. The twin UV bumps need fine-tuning on how energy is injected in these scenarios.
Therefore, this synchronized X-ray/UV re-brightening gives strong evidence for the onset emission determined by a slow coasting jet~\citep{Blandford76,Sari99}.

\begin{figure*}
	\centering
	\includegraphics[width=0.55\textwidth]{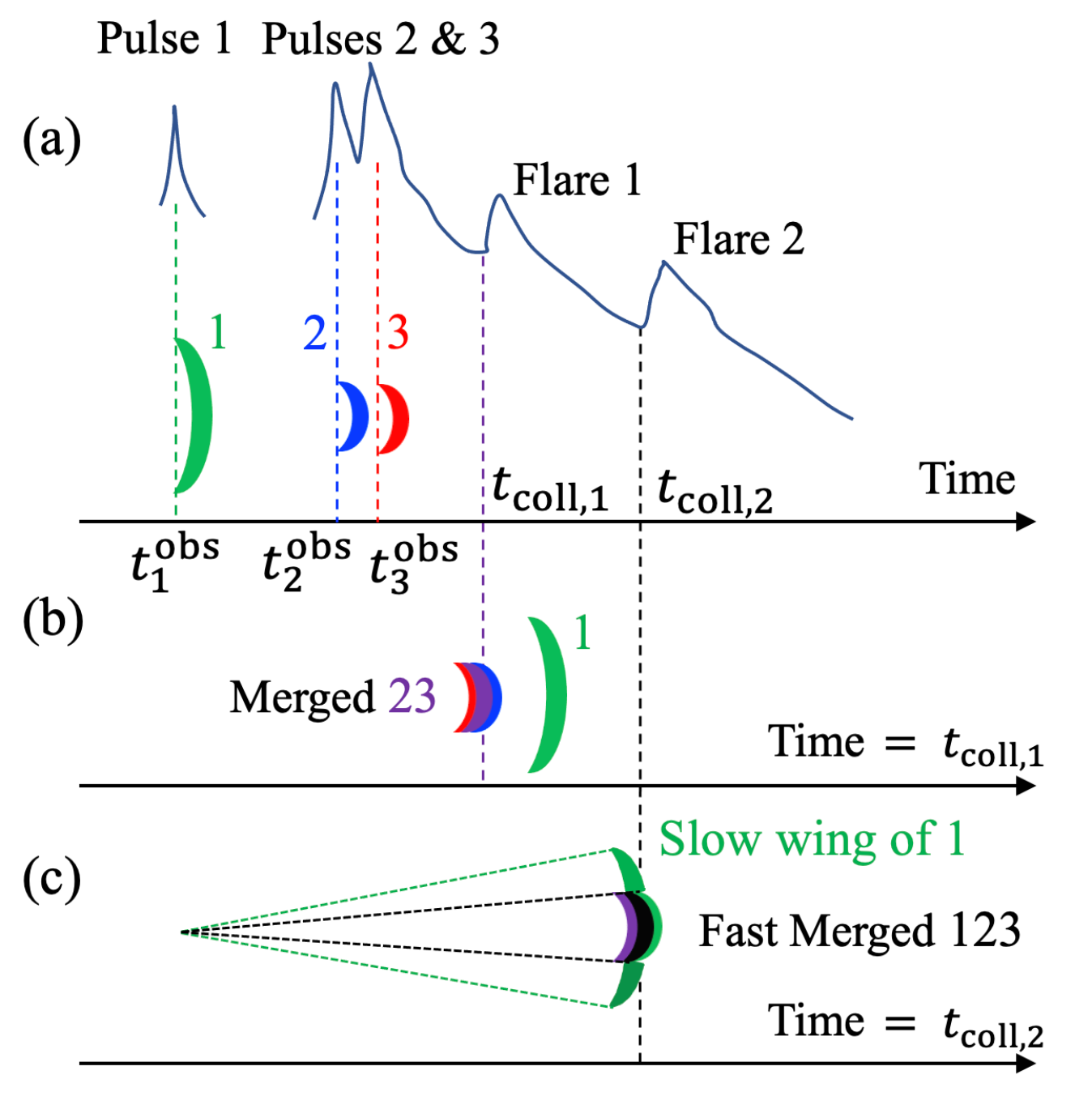}
	\caption{The schematic picture of the multiple-shock collision scenario.
	In panel (a), the colored arches represent the relativistic shells released during prompt pulses at different observational times, labeled with the increasing Arabic number (1-3).
    In panel (b), it shows that at $t_{\rm coll,1}$ in the observer frame, the 3rd shell catches up with the 2nd shell and merges as shell 23 (purple color), producing the 1st X-ray flare. In panel (c), a similar collision between shell 23 and the earlier shell 1 occurs at $t_{\rm coll,2}$ and produces the 2nd X-ray flare. The finally merged shell 123 (black color, called the fast jet) and the survival wing part of shell 1 (called the slow hollow jet) would produce the observed long-lasting afterglows when they propagate in the circumburst medium.}
    \label{fig:schematic}
\end{figure*}

\section{GRB timing}
\begin{figure*}
	\centering
	\includegraphics[width=0.8\textwidth]{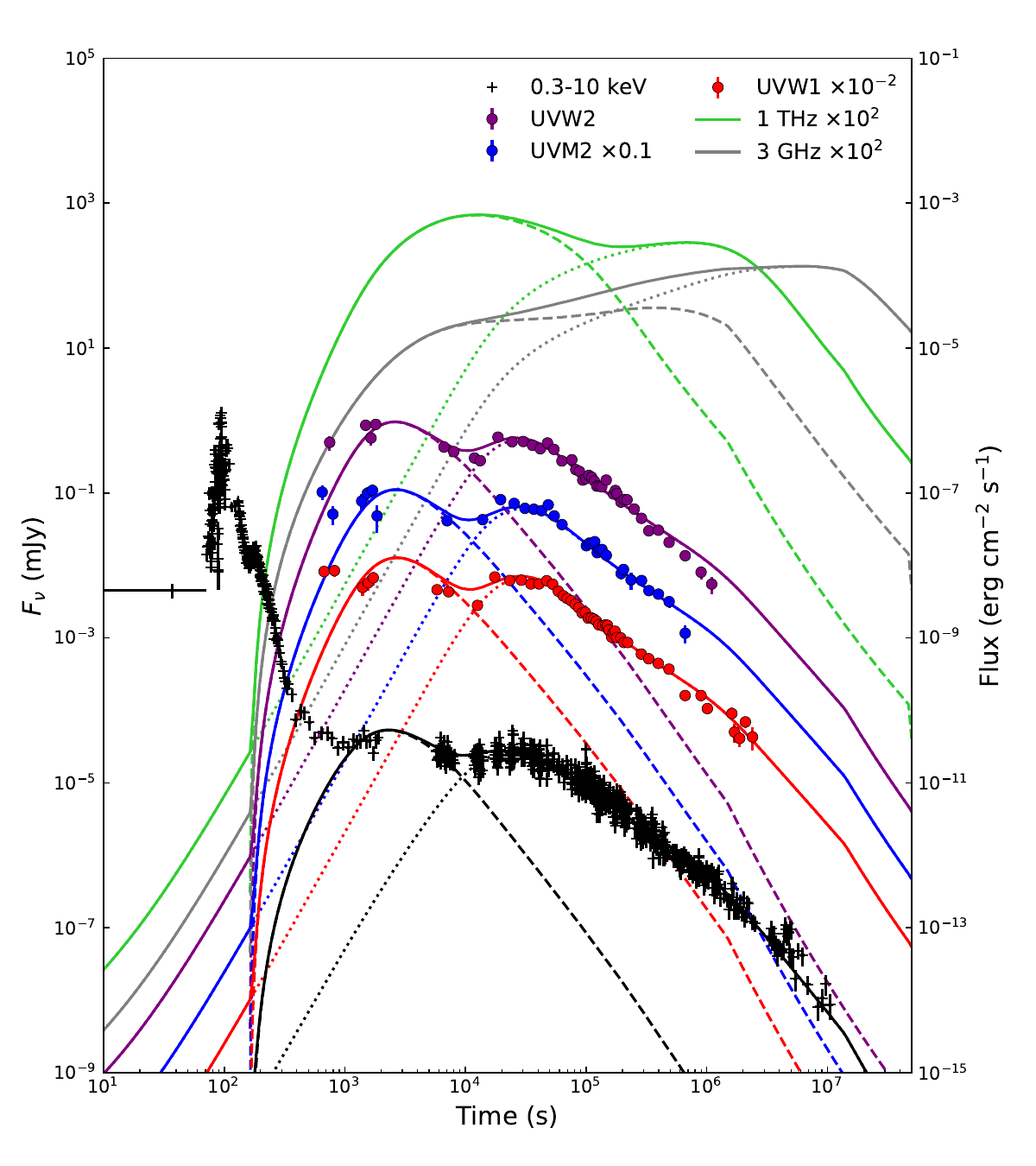}
	\caption{Fit to the multi-wavelength lightcurves of GRB 060729.
	Each colored points and lines are observational data and theoretical results in each frequency band.
	The dotted and dashed lines represent the emissions from the slow and the fast jet respectively, while the solid lines are the sum of these components.
	The predicted lightcurves at THz and 3~GHz are shown in green and grey lines respectively.
	The X-ray data are taken from the {\it Swift}/XRT website (http://www.swift.ac.uk/xrt\_curves/00221755/), and the UV data are taken from the UVOT photometry result in \cite{Grupe07}.
	} \label{fig:fit}
\end{figure*}

Given that BAT observations reveal three pulses in the main burst, it is reasonable to suppose that each pulse would drive a separate outgoing relativistic outflow from the first principle~\citep{Blandford76}.
Similar to the internal shock model originally proposed 
for the energy dissipation during the prompt emission~\citep{Rees94}, 
collisions between these outflows/shells become inevitable when their velocities are different and the fast shells are launched later.
In our framework, we first assume that the first shell is relatively slower and the late two shells are faster.
In the illustration \FIG{fig:schematic} of this multiple-shock scenario, the collision between the 2nd and the 3rd shell would make the 1st X-ray flare, and the collision between the 1st shell and the former merged shell would further result in the 2nd X-ray flare. The forming time of each shell in the observer frame is denoted as $t_{i}^{\rm obs}$ ($i = 1-3$)
and the initial bulk Lorentz factor (velocity) of the shell as $\Gamma_i$ ($\beta_i$). As the minimum variability timescale of each prompt pulse ($\le 1$~s) is much smaller than the time gaps between them ($\ge 10~$~s), we would expect that the difference in their shell-forming radii is negligible compared to the collision radii~\citep{Rees94,Golkhou15}, and two collisions occur at
\begin{eqnarray}
t_{\rm coll,1} &\simeq& t_{2}^{\rm obs} + \left[1 + (1-\beta_3) / (\beta_3 - \beta_2)\right] (t_{3}^{\rm obs} - t_{2}^{\rm obs}), \\
t_{\rm coll,2} &\simeq& t_{\rm coll,1} +  \left(1- \beta_{23} \right) / (\beta_{23} - \beta_{\rm 1,c}) \\ \nonumber
& & \times \left[\beta_{\rm 1,c} t_{3}^{\rm obs} + \beta_2 (\beta_{\rm 1,c} - \beta_3) (t_{3}^{\rm obs} - t_{2}^{\rm obs})/(\beta_3 - \beta_2) \right],
\end{eqnarray}
for the situation of $\Gamma_3 > \Gamma_2 > \Gamma_{1,c}$, where $\Gamma_{\rm 1,c}$ ($\beta_{\rm 1,c}$) is the bulk Lorentz factor (velocity) of the core part of the wide shell 1, and $\Gamma_{23}$ ($\beta_{23}$) is the bulk Lorentz factor (velocity) of the merged shell of 2 and 3.
If the half-opening angles of fast shells 2 and 3 are similar ($\theta_{\rm j,f}$) and narrower than that of the slow 1st shell ($\theta_{\rm j,s}$), after the two collisions, the final outflows could be described by two jets, an inner slow jet, and an outer slow hollow jet. The fast inner jet carries the material from three former shells and we denote its bulk Lorentz factor as $\Gamma_{\rm f}$ (i.e. $\Gamma_{123}$ in Figure~\ref{fig:schematic}).
Meanwhile, the outer hollow jet is just the original wing part of shell 1
that free of interactions and its slower bulk Lorentz factor is denoted by $\Gamma_{\rm s}$ (i.e. $\Gamma_{\rm 1,w}$). During each collision, the conservation of energy and momentum holds, i.e.,
\begin{eqnarray}
\Gamma_{23} &=& \left(\frac{\Gamma_2 m_2+\Gamma_3 m_3}{m_2/\Gamma_2 + m_3/\Gamma_3} \right)^{1/2}, \\
\Gamma_{123} &=& \Gamma_{\rm f} = \left(\frac{\Gamma_{23} (m_2+m_3) +\Gamma_{\rm 1,c} m_{\rm 1,c}}{(m_2+m_3)/\Gamma_{23} + m_{\rm 1,c}/\Gamma_{\rm 1,c}} \right)^{1/2}, 
\end{eqnarray}
if leaked radiation energy during the collision is not so significant (see below), where $m_{i}$ is the equivalent isotropic mass of each shell.
After this collisional period, the survival shocks will propagate into the circumburst environment and produce the long-term afterglow.
For a blastwave with an initial Lorentz factor of $\Gamma_{i}$ and an isotropic kinetic energy of $E_{\mathrm{K},i}$, it keeps coasting before deceleration due to sweeping up enough of the surrounding material.
The afterglow emission thus keeps rising during this coasting phase and peaks at
\begin{eqnarray}
t_{\mathrm{peak},i} &\simeq& 1.2 \times 10^3 (1 + z) \left(\frac{E_{\mathrm{K},i}}{10^{53}~\mathrm{erg}} \right)^{1/3} \\ \nonumber
& & \times
\left(\frac{n}{1~\mathrm{cm}^{-3}} \right)^{-1/3} \left( \frac{\Gamma_i}{50}\right)^{-8/3}~\mathrm{s},
\end{eqnarray}
in the standard GRB shock model~\citep{Blandford76}, where $n$ is the number density of the ambient medium experienced by the later shock.
Below, we show that the two peaks of the survival shocks are evidenced in the observed X-ray and optical afterglow (Figure~\ref{fig:fit}).

\section{Results}
The timing relations in Equations (1-5) and the standard afterglow model (Appendix \ref{append:B}) based on these shocks allow us to constrain their properties.
Following the standard Bayesian framework, we proceed to derive model parameters by numerically optimizing the likelihood function of
\begin{equation}
\ln \mathcal{L}(\vec{x}) = \ln \mathcal{L}_{\rm afterglow}(\vec{x}) + \ln \mathcal{L}_{\rm prompt}(\vec{x}), 
\end{equation}
where $\vec{x}$ is a vector containing the model parameters.
Here, the information of both the afterglow and the prompt emission are taken into account, i.e.,
\begin{equation}
\ln \mathcal{L}_{\rm afterglow}(\vec{x}) = -\frac{1}{2} \Sigma_i \left[ \frac{(F_{i,\mathrm{obs}}-F_i(\vec{x}))^2}{\sigma_i^2}  \right], 
\end{equation}
and 
\begin{equation}
\ln \mathcal{L}_{\rm promt}(\vec{x}) = 0
\end{equation}
is taken if Equations (1-4) have physical solutions, or $\ln \mathcal{L}_{\rm promt}(\vec{x}) = -\mathrm{Infinity}$ otherwise,
where $F_{i,\mathrm{obs}}$ and $\sigma_i$ are the measured afterglow flux and flux errors respectively. To account for uncertainties in the timing of the lightcurve and their impact on our results, we adopt several ranges for the critical time points with uniform probability distribution, i.e., $t_{2}^{\rm obs} \in [76, 80]$~s, $t_{3}^{\rm obs} \in [90, 92]$~s, $t_{\rm coll,1} \in [115, 118]$~s and $t_{\rm coll,2} \in [161, 165]$~s, as marked in \FIG{fig:XRT}. The shell launching times are chosen to align with the rising phases of the pulses, while the collision times correspond to the temporal turnover points in the lightcurves.
In such a synergetic model, there are fourteen parameters left free, i.e., $\Gamma_{\rm 1,c}$, $\Gamma_{\rm s}$ ($\Gamma_{\rm 1,w}$), $\Gamma_3$, $E_{\rm K,3}$, $E_{\rm K,s}$, $\theta_{\rm j,s}$, $p_{\rm s}$, $\epsilon_{\rm e,s}$, $\epsilon_{\rm B,s}$, $\theta_{\rm j,f}$, $p_{\rm f}$, $\epsilon_{\rm e,f}$, $\epsilon_{\rm B,f}$, and $n$ (see Appendix \ref{append:B} for details).
Their posteriors are derived from the standard Bayesian procedure, as shown in \TAB{tab:MCMC} and \FIG{fig:corner}. 

Our results yield that the bulk Lorentz factors of the shocks driven by the main pulses and merged shocks are $\Gamma_{\rm 1,c} = 48.98^{+3.50}_{-3.27}$, $\Gamma_2 = 62.37^{+4.35}_{-4.30}$, $\Gamma_3 = 75.86^{+5.43}_{-5.06}$, $\Gamma_{23} = 70.90^{+4.49}_{-4.58}$, and $\Gamma_{\rm f} = 57.12^{+4.06}_{-4.08}$ respectively.
The comoving thickness of the slow shell could be approximated as $\Delta \simeq 2\Gamma_{\rm 1,c} c t_{\rm coll,2}$ near $t_{\rm coll,2}$,
one may therefore expect that the 2nd flare would reach its peak luminosity when the merged shell 23 moves across shell 1 at an observational time of 
\begin{eqnarray}
T_{\rm peak} &\simeq& t_{\rm coll,2} + (1-\beta_{\rm f} \cos(\min[\theta_{\rm j,f},1/ \Gamma_{\rm f}])) \\ \nonumber
& & \times \Delta /(\beta_{23\vert\rm{1c}} c) \in [174.19, 178.61] (1\sigma)~\mathrm{s},  
\end{eqnarray}
with the derived values of $\theta_{\rm j,f} = 0.02^{+0.00}_{-0.00}$~rad and $\beta_{23\vert\rm{1c}} = 0.36^{+0.02}_{-0.02}$, where $\beta_{23\vert\rm{1c}}$ is the relative velocity between shell 23 and the core of shell 1. This perfect match with the observed peak time of the 2nd X-ray flare (see \FIG{fig:XRT}) strengthens the robustness of our results.

\FIG{fig:fit} shows that both the UVOT and XRT data are well interpreted by the best-fitting results from the posteriors.
The twin temporal peaks due to two onset emissions should also be significant in the frequency band of THz, which could help to further distinguish our multiple-shock scenario from other possible scenarios for similar bursts in the future.
The realistic jets should be structured~\citep{Tchekhovskoy08,Lazzati12,Geng16b,Geng19,Gottlieb20}, especially for the wide wing part of the 1st shell. In fact, the required $\Gamma_{\rm 1,c} > \Gamma_{\rm 1,w}$ in our calculations supports the 1st wide jet is initially structured itself.
However, the observational data are not insufficient to constrain relevant structure parameters strictly, especially when the jet axis is aligned to our line of sight~\citep{Xu23,OConnor24}.
Therefore, we use a top-hat jet approximation in current calculations.
The additional long-term monitoring in THz and radio in similar bursts could, in principle help to constrain the jet structure. 

The circumburst environment in our scenario is assumed to be a homogeneous interstellar medium. The coasting jet in the wind-like medium has been proposed to interpret X-ray plateaus in some GRBs~\citep{Begue22}, where the predicted temporal behavior is $F_{X} \propto t_{\rm obs}^{(2-p)/2}$. Since the statistical results show evidence of brightening behavior in both X-ray and optical bands around $\sim 10^4$~s and $p \ge 2$ is usually expected, it would not be easy to match the observations for the case of wind environment.

The geometry configuration of the jets for afterglows here seems similar to the traditional two-component jet scenario used for some GRBs in the literature.
However, we should address the fact that our scenario is substantially different from the two-component jet scenario.  
Our multiple-shock scenario is rooted in the evolving nature of the outflow with time or along the radial direction equivalently. In contrast, the traditional two-component jet scenario focuses on the jet's original structure when it is produced from the central engine, and the potential interactions between these jets during the early phase are ignored.

\cite{Moss23} have proposed that the multiple optical fluctuations in the decaying afterglow lightcurve of GRB 030329 are produced via refreshed shocks. In their scenario, an initially fast shell decelerates and is subsequently overtaken by a cluster of slower shells beyond $\sim~1$ day, representing one possible realization within the multiple shock framework.
The explicit incorporation of jet geometry (narrow versus wide, Figure~\ref{fig:schematic}), and the direct physical connection between jet dynamics and both prompt emission pulses and X-ray flares make our scenario different from previous works.
Furthermore, the collisions in GRB 060729 occurred much earlier than the shock deceleration time, which makes the timing analyses more straightforward and allows us to perform robust afterglow modeling. 

Equation (5) alone has been widely used as a traditional method  to estimate the bulk Lorentz factors of GRB jets~\citep[e.g.,][]{Liang10,Liliang12,Liang13,Ghirlanda18}. While this approach is effective for bursts exhibiting clear onset signatures and power-law decaying afterglows, it relies solely on this single relation without incorporating additional constraints. To assess its validity, we perform a comparative calculation using two jets with an assumed geometry same as Figure~\ref{fig:schematic} but without applying the timing constraints from Equations (1-4). The resulting bulk Lorentz factor of $25.70^{+2.48}_{-1.72}$ and $53.70^{+2.53}_{-2.42}$ for the slow and fast jet respectively, are consistent with the results in Table~\ref{tab:MCMC} under the full consideration of timing relations. However, this simplified approach fails to explain the origin of the jet geometry or the X-ray flares, and it is not directly connected with the ``plateau'' feature at first glance. Our paradigm, in contrast, is particularly efficient for bursts with multiple prompt pulses or complex afterglow features that deviate from standard behavior~\citep{WangXG2015}. By integrating timing relations, it provides a more coherent picture bridging the prompt and afterglow phases.

\section{Discussion}
Our results indicate that the initial bulk Lorentz factors of the shocks driven by the main pulses in GRB 060729 are substantially lower than the typical values of several hundred observed in other GRBs~\citep[e.g.,][]{Liang10,Yi17,Ghirlanda18}.
The relatively slow nature of these jets implies a higher baryon loading compared to typical GRB jets, which aligns with an accretion-driven origin~\citep{Lei13,Liu17}.
Nevertheless, the increasing sequence of Lorentz factors, $\Gamma_{1,c} < \Gamma_2 < \Gamma_3$ supports a progressive reduction in baryon loading along with central engine activity.
This trend is consistent with the scenario of a magnetically arrested disk, where the magnetic field plays a dominant role in regulating the accretion process~\citep{Tchekhovskoy08,Gottlieb24}.
Further investigations into the origin of such slow jets and their potential connections to the types of central compact objects (e.g., neutron stars or black holes) are encouraged.

The jet launching sequence has a significant impact on their subsequent evolution. In GRB 240529A, where a fast jet is launched ahead of slower components, collisions between the jets are avoided. In contrast, for GRB 060729, the early-launched slow and wide jet acts as a ``speed bump'' for any later faster jets, effectively slowing down the velocity of the merged/survival shocks. 
This mechanism indicates a potentially large sample of slow shocks resulting from such ordered collisions, in addition to the intrinsically slow shock. 

The X-ray flares of GRB 060729 here are naturally interpreted as the outcomes of collisions between the launched shocks, which also aligns with the relativistic shell models for X-ray flares~\citep{Uhm15,Geng17,Geng18c}.
By combining the constrained kinetic energy and velocity of these shocks, we can estimate the radiation efficiency of the collision process, defined as $\eta = E_{\rm rad}/(E_{\rm rad}+E_{\rm K})$, where $E_{\rm rad}$ is the observed radiation energy of the main pulses or X-ray flares.
\FIG{fig:eff} shows the relationship between $\eta$ and the relative velocity between the shocks for the prompt pulses and X-ray flares.
The radiation efficiency of the X-ray flares ($\le 10^{-2}$) is lower than that of the prompt pulses.
This is consistent with theoretical expectations, as internal shocks or magnetic reconnection occurring at smaller radii should result in stronger dissipation than gradual collisions at larger radii.
More burst samples with diverse shock combinations could help to elucidate the radiation efficiency of different collision scenarios and its underlying correlation with the relative velocity or other relevant parameters.

\begin{table}
\caption{Parameters values used in the modeling of the afterglow of GRB 060729 inferred from the marginal posterior distributions shown in \FIG{fig:corner}.}
\label{tab:MCMC}
\centering
\begin{tabular}{c c c}
\hline\hline
Parameters & Range of Priors & Posteriors ($1\sigma$) \\
\hline
$\Gamma_{\rm 1,c}$   & [10, 200] & $48.98^{+3.50}_{-3.27}$    \\
$\Gamma_{\rm 1,w}$   & [10, 200] & $23.50^{+1.68}_{-1.56}$    \\
$\Gamma_{3}$         & [10, 200] & $75.86^{+5.43}_{-5.06}$    \\
$E_{\rm K,s}$~($10^{52}$~erg) &  [0.1, 10]   &  $4.37^{+0.10}_{-0.10}$ \\
$E_{\rm K,3}$~($10^{52}$~erg) &  [0.1, 10]   &  $2.45^{+4.15}_{-1.17}$ \\
$\theta_{\rm j,s}$ (rad)      &  [0.1, 0.3]  &  $0.25^{+0.02}_{-0.02}$ \\
$p_{\rm s}$                   &  [2.05,2.5]  &  $2.20^{+0.01}_{-0.01}$ \\
$\epsilon_{\rm e,s}$          &  [$10^{-3}$,0.5] & $0.49^{+0.00}_{-0.01}$ \\
$\epsilon_{\rm B,s}$          &  [$10^{-4}$,0.1] & $0.02^{+0.00}_{-0.00}$ \\
$\theta_{\rm j,f}$ (rad)      &  [0.01, 0.1] & $0.02^{+0.00}_{-0.00}$  \\
$p_{\rm f}$                   &  [2.05,2.5]  & $2.30^{+0.08}_{-0.07}$  \\
$\epsilon_{\rm e,f}$          &  [$10^{-3}$,0.5] & $0.04^{+0.01}_{-0.01}$ \\
$\epsilon_{\rm B,f}$          &  [$10^{-4}$,0.1] & $0.03^{+0.01}_{-0.01}$ \\
$n$~(cm$^{-3}$)               &  [$10^{-2}$, 10] & $0.17^{+0.06}_{-0.05}$ \\
\hline
\end{tabular}
\end{table}

\section{Conclusions} \label{sec:conclusion}
Our comprehensive analysis of the remarkable GRB 060729 uncovers properties that echo the double-yolk burst GRB 240529A detected recently. Employing the newly developed diagram of the synergy analyzing method, we demonstrate that both the X-ray flares and the X-ray/UV afterglows 
can be naturally explained by the scenario involving three interacting shocks with an initial bulk Lorentz factor of only several tens. This approach provides a unified interpretation of the observed phenomena, bridging the prompt and afterglow phases.

Our work brings several novel insights into the GRB field.
First, the synergy analysis of the prompt and afterglow data pave a new way to probe shock physics. By leveraging the timing information of prompt pulses and X-ray flares, we could now measure the velocities of these shocks with high confidence.
Moreover, we propose that the X-ray plateau, a commonly observed feature, may arise from the superposition of two or several onset emission components.
Alternatively, it could result from the combination of one onset emission component with a decaying component, as seen in GRB 240529A. This highlights the importance of conducting detailed analyses of both the overall morphology and the substructure of X-ray lightcurves in future studies.
Finally, our findings underscore the critical role of early shock interactions in shaping the final geometry of the outflow propagating in the circumburst environment, particularly in GRBs exhibiting multiple pulses or emission episodes during the prompt phase.

The discoveries from GRB 060729 and GRB 240529A suggest that a significant population of slow jets ($\Gamma \le 100$) may have been missed in previous studies. This oversight could stem from several factors, including incomplete X-ray data sampling during the critical period of $~10^2 -10^5$~s post-burst, the lack of simultaneous multi-wavelength observations, the dominance of a single bright jet component in multi-jet systems, and the inherent complexity of analyzing intricate prompt emission phases.
Despite these challenges, a systematic search for similar bursts harboring slow jets is imperative.
Such efforts will shed light on the diversity of central engine activities and their associated outflows.

The success of the {\it Swift}~\citep{Gehrels04}, Fermi satellite~\citep{Meegan09,Atwood09} has greatly advanced our understanding of GRBs.
The Space Variable Objects Monitor~\citep{Wei16} focusing on GRBs and the Einstein Probe satellite~\citep{Yuan22,SunH24} dedicated to time-domain high-energy astrophysics have been launched and successfully put into operation.
The incoming fruitful data along with our synergy analyzing method will allow us to delve into the GRB jet/shock physics.

\vspace{\baselineskip}
We appreciate valuable comments and suggestions from the anonymous referee.
This study is partially supported by the National Natural Science Foundation of China (Grant Nos. 12273113, 12321003, 12393812, 12393813 and 12233002), the National SKA Program of China (Grant Nos. 2022SKA0130100 and 2020SKA0120300), the Strategic Priority Research Program of the Chinese Academy of Sciences (Grant No. XDB0550400), the National Key R\&D Program of China (2021YFA0718500) and the FAST Special Program (NSFC 12041301). Jin-Jun Geng acknowledges support from the Youth Innovation Promotion Association (2023331). Yong-Feng Huang and Tao An also acknowledge the support from the Xinjiang Tianchi Program.

\begin{figure*}[ht]
	\centering
	\includegraphics[width=\textwidth]{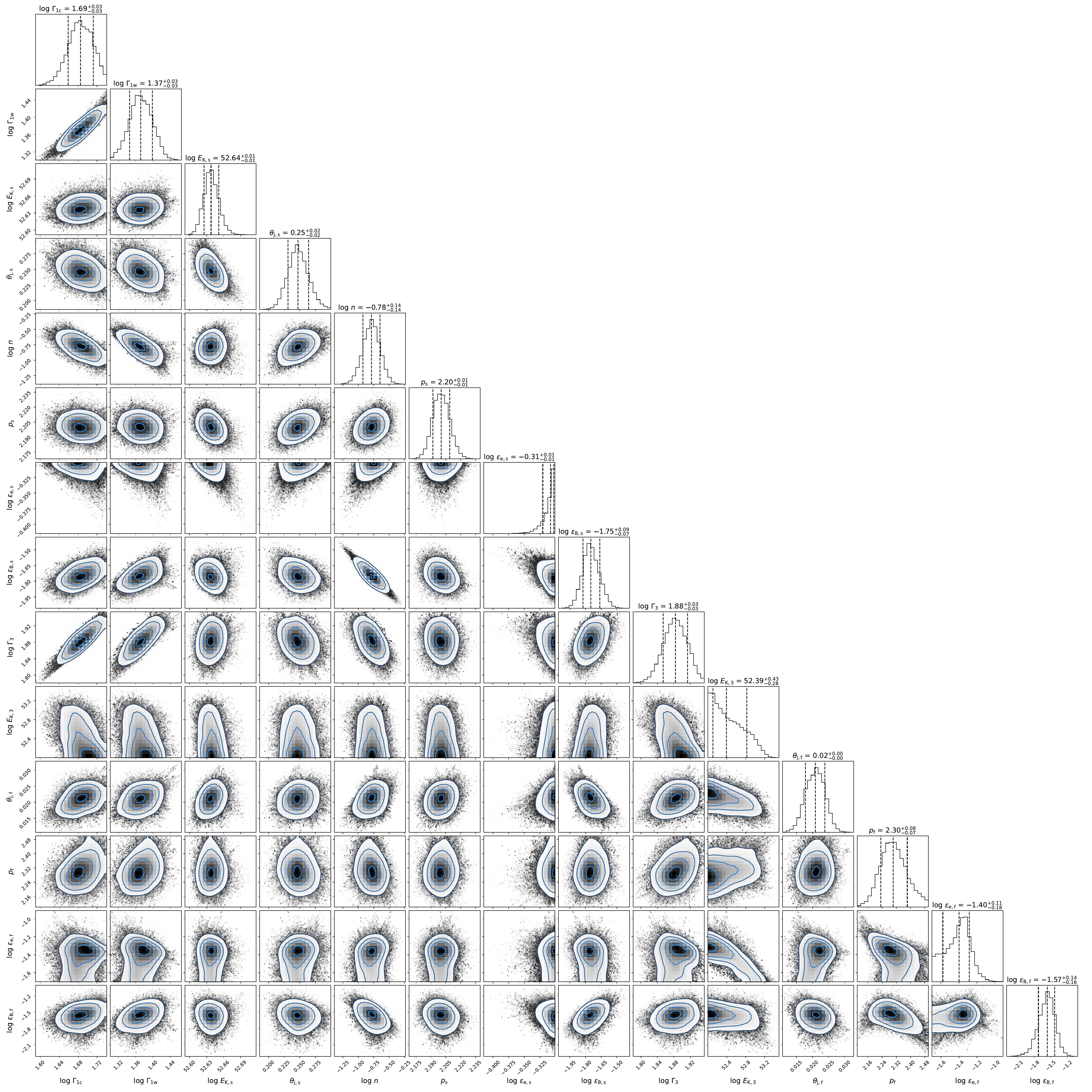}
	\caption{The corner plot of the fitting result in \TAB{tab:MCMC}.
	The corner plot shows all the one and two-dimensional projections of the posterior probability distributions of fourteen parameters used in the model. The 1-dimensional histograms are marginal posterior distributions of these parameters. The vertical dashed lines indicate the 16th, 50th, and 84th percentiles of the samples, respectively, which are labeled on the top of each histogram.} \label{fig:corner}
\end{figure*}

\begin{figure*}
	\centering
	\includegraphics[width=0.8\textwidth]{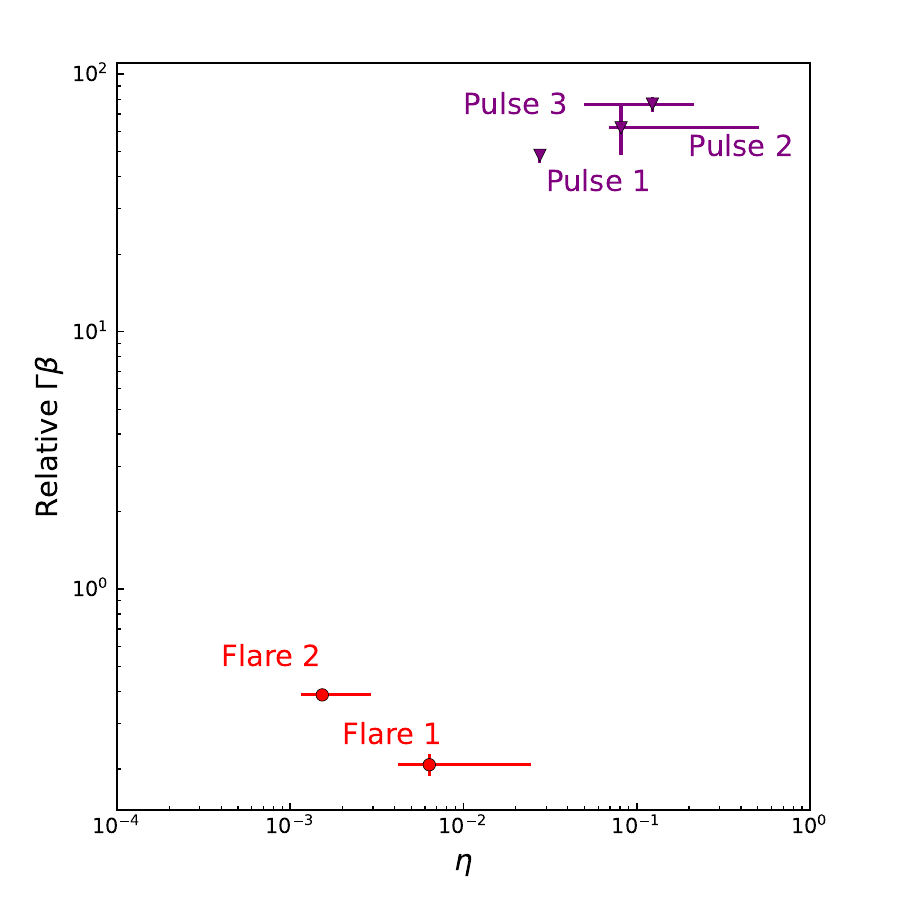}
	\caption{The relationship of radiation efficiency ($\eta$) versus the relative velocity ($\Gamma \beta$) between shocks. The red and purple points present the flares produced by the collision and three prompt pulses respectively. The radiation energy of pulses and flares are taken from Table 1 of \cite{Grupe07}.
	Note that the bulk velocity is adopted as the upper limit of the relative velocity between the internal shocks for three prompt pulses.
	} \label{fig:eff}
\end{figure*}

\clearpage
\bibliography{main.bib}

\appendix
\section{The Statistics} \label{append:A}
\begin{figure*}
	\centering
	\includegraphics[width=0.8\textwidth]{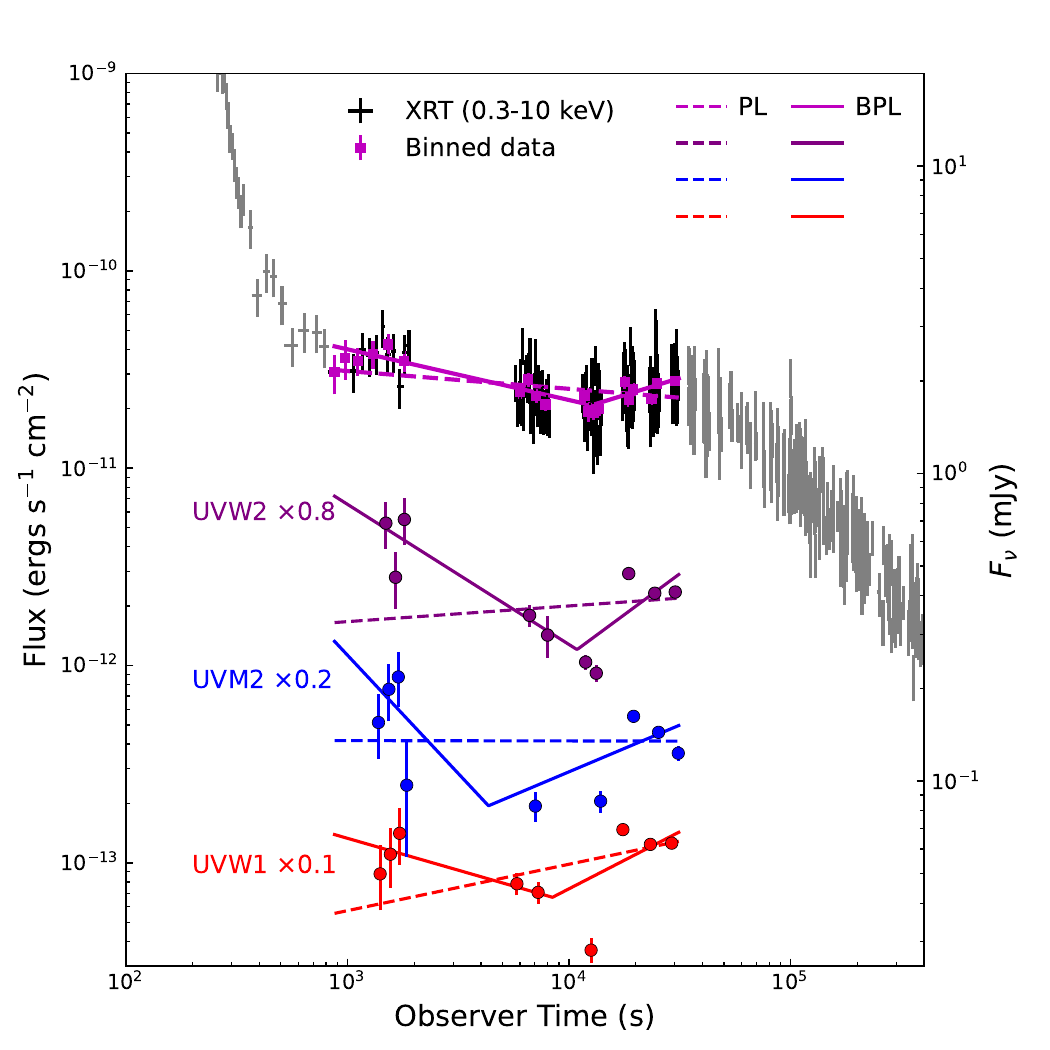}
	\caption{Analytical fit to the X-ray and optical data in the time range of $\simeq [10^3, 3 \times 10^4]$~s.
	The segment of the original X-ray data (black) is binned into twenty points (magenta). For each band marked with different color, the data is fitted by the power law (PL, dashed line) and broken power law (BPL, solid line), respectively, whose fitting goodness and temporal indices are listed in Table~\ref{tab:fit_bin}.
	} \label{fig:rebin}
\end{figure*}

Here, we perform the data binning to verify the dependence of the model preference on the number of X-ray data.
The K-Means clustering algorithm~\citep{MacQueen67} is adopted to identify the natural groupings of data points in the logarithmic timeline, from which the data are binned adaptively based on their temporal density, ensuring statistically robust representation while preserving critical features of the lightcurve.
For the $k$th cluster, its time centroid is determined by the geometric mean of involving points weighted by exposure time, and the binned average flux ($F_{\mathrm{bin},k}$) and error ($\sigma_{\mathrm{bin},k}$) could be estimated as
\begin{equation}
F_{\mathrm{bin},k} = \frac{\Sigma_i F_{i} \Delta t_i}{\Sigma_i \Delta t_i}, ~~~\sigma_{\mathrm{bin},k} = \frac{\sqrt{\Sigma_i \sigma_i^2 \Delta t_i^2}}{\Sigma \Delta t_i},
\end{equation}
where $F_i$ ($\pm \sigma_i$) and $\Delta t_i$ are the flux (error) and the exposure time of $i$th data point within each cluster. 
Twenty binned data points are derived with this method (see magenta points in Figure~\ref{fig:rebin}).
The resulting difference in AIC and BIC for power-law (PL) fit and broken power-law (BPL) fit suggests that the BPL is preferred at a $>6\sigma$ level (Table~\ref{tab:fit_bin}), consistent with the conclusion in Section 2.

Meanwhile, the optical data within the similar time segment ($\simeq [10^3, 3 \times 10^4]$~s) are fitted with the two functions (Table~\ref{tab:fit_bin} and Figure~\ref{fig:rebin}). The BPL is also found to be preferred at a $>6\sigma$ level for each optical band. The PL indices are even positive for the UVW2 and UVW1 data. This further supports that a new emission component is emerging around $10^4$~s in both the X-ray and optical bands.

\section{Afterglow Model} \label{append:B}
The pair of two onset emissions gives the ratio between the bulk Lorentz factors of two jets as
\begin{equation}
\frac{\Gamma_{\rm f}}{\Gamma_{\rm s}} =
\left(\frac{E_{\mathrm{K,s}}}{E_{\mathrm{K,f}}} \right)^{-1/8}
\left(\frac{t_{\rm peak,s}}{t_{\rm peak,f}}\right)^{3/8},
\label{eq:ratio_G}
\end{equation}
which is insensitive to the difference in the kinetic energy. 
The afterglow lightcurve indicates that $t_{\rm peak,s} \simeq 2 \times 10^4$~s and $t_{\rm peak,f} \simeq 2 \times 10^3$~s, hence we obtain $\Gamma_{\rm f} / \Gamma_{\rm s} \simeq 2.4$. Note that Equation (\ref{eq:ratio_G}) is only valid for two on-axis viewed jets, an off-axis-viewing correction should be accounted for the hollow slow jet in more accurate estimations.

The afterglow emission is considered to be synchrotron radiation of electrons accelerated by relativistic shocks. The synchrotron spectra are characterized by two critical frequencies, the typical frequency of $\nu_{\rm m} \propto \Gamma B^{\prime} \gamma_{\rm m}^2$
and the cooling frequency of $\nu_{\rm c} \propto \Gamma B^{\prime} \gamma_{\rm c}^2$, together with the peak flux density $F_{\nu,\mathrm{max}} \propto \Gamma N_{\rm e} B^{\prime}$ by a total of $N_{\rm e}$ electrons, where $\gamma_{\rm m}$ and $\gamma_{\rm c}$ are the minimum Lorentz factor and the cooling Lorentz factor of the electron spectrum, and $B^{\prime}$ is the comoving frame magnetic field post the shock.
After introducing the microphysical shock parameters like equipartition parameters shocked electrons and magnetic field ($\epsilon_{\rm e}$, $\epsilon_{\rm B}$), and the electron spectral index ($p$), we have $\nu_{\rm m} \simeq 1.3 \times 10^{14} (\Gamma / 50)^{7/2} n_0^{1/2} \epsilon_{\rm e,-1}^2 \epsilon_{\rm B,-2}^{1/2}~\mathrm{Hz}$, $\nu_{\rm c} \simeq 1.3 \times 10^{19} (\Gamma / 50)^{-5/2} n_0^{-3/2} \epsilon_{\rm B,-2}^{-3/2} (t_{\rm obs} / t_{\rm peak})^{-2}~\mathrm{Hz}$, and $F_{\nu,\mathrm{max}} \simeq 5.4 (\Gamma / 50)^{15/2} n_0^{3/2} \epsilon_{\rm B,-2}^{1/2} (t_{\rm obs} / t_{\rm peak})^{3}~\mathrm{mJy}$
for $t_{\rm obs} \le t_{\rm peak}$, where a typical value of $p = 2.3$ is adopted for simplicity.
During the rising phase, the spectrum is roughly in the $\nu_{\rm m} < \nu_{\rm UV} < \nu_{\rm X} < \nu_{\rm c}$ regime, and the flux density in UV band is hence $F_{\nu,\mathrm{UV}} \propto F_{\nu,\mathrm{max}} (\nu_{\rm UV} / \nu_{\rm m})^{-(p-1)/2}$. If we further assume that $p$ does not differ significantly for the two jets here, the peak time and the peak flux of the twin UV bumps give
\begin{equation}
\frac{E_{\rm K,f}}{E_{\rm K,s}} =
\left(\frac{F_{\mathrm{UV,peak,f}}}{F_{\mathrm{UV,peak,s}}} \right)^{32/(41-7p)}
\left(\frac{t_{\rm peak,f}}{t_{\rm peak,s}}\right)^{(27-21p)/(41-7p)}
\left(\frac{\epsilon_{\rm e,f}}{\epsilon_{\rm e,s}}\right)^{32(1-p)/(41-7p)}
\left(\frac{\epsilon_{\rm B,f}}{\epsilon_{\rm B,s}}\right)^{-8(p+1)/(41-7p)},
\end{equation}
which provides useful relations for inferring the shock parameters analytically. Again, since the UV flux density from a realistic jet is sensitive to other geometry factors including the half-opening angle of the jet and the viewing angle, detailed numerical calculations are essential to explore the reasonable values of these parameters.

In our numerical calculations, jet dynamics during the afterglow phase are described by a set of dynamical equations of the relativistic blastwave~\citep{Huang99,Huang00,Peer12} incorporating the jet sideways expansion~\citep{Granot12}. The continuity equation of electrons accelerated by the shock is strictly solved to obtain the time-dependent electron spectra~\citep{Geng18b}. Specifically, the ratio of the non-thermal electron energy to the whole shocked electron energy is taken as a typical value of $0.5$~\citep{Giannios09,Gao24}.
Denoting the initial half-opening angles of the fast and the slow jets as $\theta_{\rm j,f}$ and $\theta_{\rm j,s}$,
the multi-wavelength afterglow emissions are derived by integrating the equal-arrival-time surface~\citep{Granot99}, i.e., $\int_0^{\theta_{\rm j,f}} ... d \theta$ for the inner one and $\int_{\theta_{\rm j,s}}^{\theta_{\rm j,f}} ... d \theta$ for the hallow one.
By embedding our numerical module into the Markov chain Monte Carlo Ensemble sampler called emcee~\citep{Foreman-Mackey13}, the posteriors of our model parameters are obtained as shown in \TAB{tab:MCMC} and \FIG{fig:corner}.

\begin{table}[htbp]
\centering
\caption{Fitting results of the X-ray and optical data with analytical functions.}
\label{tab:fit_bin}
\begin{tabular}{cccccccc}
\toprule
\multirow{2}{*}{Band} & \multirow{2}{*}{Model} & \multicolumn{3}{c}{Temporal Index} & \multirow{2}{*}{$\chi^2$/dof}
& \multirow{2}{*}{AIC} & \multirow{2}{*}{BIC} \\
\cmidrule(lr){3-5}
 &  & $\alpha$ & $\alpha_1$ & $\alpha_2$ &  &  & \\
\midrule
\multirow{2}{*}{X-ray} 
 & PL & $-0.08 \pm 0.02$ & -- & -- & 219.49/138 & 223.49 & 229.38 \\
 & BPL & -- & $-0.26 \pm 0.03$ & $0.35 \pm 0.08$ & 155.33/136 & 163.33 & 175.10 \\
\midrule
\multirow{2}{*}{X-ray (binned)} 
 & PL & $-0.09 \pm 0.02$ & -- & -- & 70.56/18 & 74.56 & 76.56 \\
 & BPL & -- & $-0.26 \pm 0.04$ & $0.34 \pm 0.09$ & 22.56/16 & 30.56 & 34.54 \\
\midrule
\multirow{2}{*}{UVW2} 
 & PL & $0.05 \pm 0.03$ & -- & -- & 256.67/8 & 260.67 & 261.27 \\
 & BPL & -- & $-0.45 \pm 0.08 $ & $0.53 \pm 0.06$ & 93.10/6 & 101.10 & 102.31 \\
\midrule
\multirow{2}{*}{UVM2} 
 & PL & $0.00 \pm 0.04$ & -- & -- & 104.60/7 & 108.60 & 108.99 \\
 & BPL & -- & $-0.77 \pm 0.45$ & $0.31 \pm 0.07$ & 62.69/5 & 70.69 & 71.48 \\
\midrule
\multirow{2}{*}{UVW1} 
 & PL & $0.15 \pm 0.03$ & -- & -- & 133.842/7 & 137.84  & 138.24 \\
 & BPL & -- & $-0.20 \pm 0.14$ & $0.38 \pm 0.10$ & 84.68/5 & 92.68 & 93.46 \\
\bottomrule
\end{tabular}
\end{table}

\end{document}